\documentclass{PoS}
\makeatletter
\def\@biblabel#1{}
\makeatother

\title{Statistical methods for the analysis of rotation measure grids in large scale structures in the SKA era}

\ShortTitle{Statistical methods for the analysis of rotation measure grids in the SKA era}

\author{\speaker{Vacca V.$^a$}, Oppermann N.$^b$, En{\ss}lin T.$^{a, c}$, Selig
  M.$^{a, c}$, Junklewitz H.$^d$, Greiner M.$^{a, c}$, Jasche J.$^e$, Hales,
  C.\ A.$^{f, g}$, Reinecke M.$^a$,  Carretti E.$^h$, Feretti L.$^i$,  Ferrari
  C.$^j$, Giovannini G.$^{i, k}$, Govoni F.$^l$, Horellou C.$^m$, Ideguchi
  S.$^n$, Johnston-Hollitt M.$^o$,  Murgia M.$^l$, Paladino R.$^{i, k}$, Pizzo R.\ F.$^p$,  Scaife A. $^q$\\
	$^a$ Max Planck Institute for Astrophysics, Karl-Schwarzschild-Str. 1,  85748 Garching, Germany\\
        $^b$ Canadian Institute for Theoretical Astrophysics, University of Toronto, 60 St. George Street, Toronto ON, M5S 3H8, Canada\\
        $^c$ Ludwig-Maximilians - Universit{\"a}t M{\"u}nchen, Geschwister-Scholl-Platz 1, 80539, München, Germany\\
        $^d$ Argelander-Institut f\"ur Astronomie, Auf dem H\"ugel 71, 52121 Bonn, Germany\\
        $^e$ CNRS, UMR 7095, Institut d'Astrophysique de Paris, 98 bis,
  boulevard Arago, F-75014 Paris, France\\
        $^f$ Jansky Fellow of the National Radio Astronomy Observatory\\
        $^g$ National Radio Astronomy Observatory, PO Box 0, Socorro, NM 87801, USA\\
        $^h$ CSIRO Astronomy and Space Science PO Box 276. Parkes 2870\\
        $^i$ INAF-Istituto di Radioastronomia, Via P.Gobetti 101, 40129 Bologna, Italy\\
        $^j$ Universit$\acute{e}$ de Nice Sophia Antipolis, CNRS, Observatoire de la C$\hat{o}$te d'Azur, Laboratoire Cassiop$\acute{e}$e, Nice, France\\
        $^k$ Department of Physics and Astronomy, University of Bologna, V.le Berti Pichat 6/2, 40127 Bologna, Italy\\
        $^l$ INAF-Osservatorio Astronomico di Cagliari, Strada 54, Loc. Poggio dei Pini, 09012 Capoterra (Ca), Italy\\
        $^m$ Department of Earth and Space Sciences, Chalmers University of Technology, Onsala Space Observatory, SE-439 92 Onsala, Sweden\\
        $^n$ University of Kumamoto, 2-39-1, Kurokami, Kumamoto 860-8555, Japan\\
        $^o$ School of Chemical \& Physical Sciences, Victoria University of Wellington, PO Box 600, Wellington 6014, New Zealand\\
        $^p$ ASTRON, Postbus 2, 7990 AA, Dwingeloo, The Netherlands\\
        $^q$ University of Southampton Highfield, Southampton SO17 1BJ, U.K.\\
        E-mail: \email{vvacca@mpa-garching.mpg.de}}

\abstract{To better understand the origin and
  properties of cosmological magnetic fields, a detailed knowledge of magnetic
  fields in the large-scale structure of the Universe (galaxy clusters,
  filaments) is crucial. We propose a new statistical approach to study
  magnetic fields on large scales with the rotation measure grid data that 
  will be obtained with the new generation of radio interferometers.}

\FullConference{
Advancing Astrophysics with the Square Kilometre Array\\
June 8-13, 2014\\
Giardini Naxos, Italy}

\begin{document}

\section{Introduction}

Magnetic fields have been detected in planets, stars, galaxies, and clusters
of galaxies.  Intracluster magnetic fields are characterized by a strength
of $\sim \mu$G and fluctuations on scales from a few kpc to $\sim100$\,kpc (e.g.,
En{\ss}lin \& Vogt 2003, Vogt \& En{\ss}lin 2003, Vogt \& En{\ss}lin 2005, Murgia
et al. 2004, Govoni et al.\ 2006, Guidetti et al.\ 2008, Bonafede et al.\ 2009, 
Bonafede et al.\ 2010, Vacca et al.\ 2010, Vacca et al.\ 2012).  On
larger scales, along galaxy filaments as well as in voids and sheets, a firm
detection has not yet been possible.  Cosmological magneto-hydro-dynamical
simulations indicate that magnetic fields are present in intergalactic
low-density environments (e.g., Dolag et al.\ 2008), while $\gamma$-ray
observations suggest their presence in voids where a lower limit of
$\sim1$\,fG has been derived (see Beck A. et al.\ 2013 and references
therein).  For more details of our current observational understanding of 
magnetic fields in low-density regions, see also the chapter by Giovannini et
al.\ (2015) in these proceedings.

Magnetic fields can be studied using a variety of techniques,
among which rotation measure (RM) observations are a powerful tool (e.g., Carilli
\& Taylor 2002, Govoni et al.\ 2004).  The amount of Faraday rotation measured
from radio observations along a given line of sight is the result of the
contribution from the Milky Way, the radio source itself and other
extragalactic environments.  The distribution
of matter in large-scale structures can be classified into four types 
(following Hahn et al.\ 2007, using a
criterion of local stability of test particle orbits): voids,
sheets, filaments and galaxy clusters.  We will hereafter call these four
types, as well as the sources responsible for the radio emission,
\emph{environments}.

Disentangling the contributions from different extragalactic environments is
essential in order to study magnetic fields in large-scale structures, and
consequently infer information about cosmic magnetism (e.g., Beck R. et
al.\ 2013, Gaensler et al.\ 2004).  However, this is a non-trivial task
that demands a statistical approach.  The first statistical studies to prove
the existence of Faraday rotation from large-scale extragalactic environments
(galaxy clusters) were published by Lawler \& Dennison (1982) and Vall{\'e}e
et al.\ (1986). Later, an important statistical RM analysis on
intracluster magnetic fields was performed by Clarke et al.\ (2001, 2004). Relevant
contributions regarding intergalactic magnetic fields on even larger scales
were done by Kolatt (1998), Stasyszyn et al.\ (2010), and Akahori et
al.\ (2014).

A statistical theory for Bayesian inference on spatially distributed signals,
\emph{Information Field Theory}, was developed by En{\ss}lin et
al.\ (2009). Within the framework of this theory, Oppermann et al.\ (2012,
2014) reconstructed the Galactic Faraday rotation foreground and 
estimated the extragalactic contribution. In this chapter we take a step
forward, with a new algorithm, able to statistically separate the different
extragalactic contributions, by combining RMs with information
about the large-scale structure of the cosmic web.  This approach will make it possible
to study magnetic fields on large scales with the RM grid data
that will be obtained with the new generation of radio interferometers (see
the chapter by Johnston-Hollitt et al.\ 2015 in these proceedings).

\section{General description}

The probability density distribution of the extragalactic Faraday rotation
measure along a given line of sight $i$ can be described with a Gaussian of
mean $\langle\phi_{{\rm e},i}\rangle$ and variance $\langle\phi_{{\rm
    e},i}^2\rangle$
\begin{eqnarray}
\langle\phi_{{\rm e},i}\rangle&=&0\nonumber\\
\langle\phi_{{\rm e},i}^{2}\rangle&=&a_{0}^{2}\int_{0}^{x_{i}}\int_{0}^{x_{i}}\mathrm{d}x\mathrm{d}x^{\prime}\frac{n_{\rm e}(x)n_{\rm e}(x^{\prime})\langle B_{x}(x)B_{x}(x^{\prime})\rangle}{(1+z(x))^2(1+z(x^{\prime}))^2}
\end{eqnarray}
where $a_0$ is a constant, $x_{i}$ is the proper distance of the source, $z$
is the redshift, $n_{\rm e}$ is the electron density, and
$B_{x}$ is the component of the magnetic field along the line of sight.  We
can define a length scale $\lambda_{x}$, so that the auto-correlation tensor
of the magnetic field $\langle B_{x}(x)B_{x}(x^{\prime})\rangle$ on average
gives significant contributions for $|x-x^{\prime}| < \lambda_{x}$, but not
for $|x-x^{\prime}| \gg \lambda_{x}$.  The strength of the magnetic field, its
correlation length, and the electron density have a different value in each
environment $j$, i.e., in voids, sheets, filaments, galaxy clusters, and the
in radio source itself.  If we define
\begin{equation}
\chi_{x}=a_0^2\,n_{\rm e}(x)^2\langle B_{x}(x)^2\rangle\lambda_{x},
\label{chidef}
\end{equation}
which we assume to depend only on the environment, so that
$\chi_{x}=\chi_{j(x)}$, and introduce $l_{ij}$ as the length of the line of
sight $i$ through each environment $j$, the overall variance in Faraday
rotation can be written as
\begin{equation}
\langle\phi_{{\rm e}, i}^2\rangle\approx a_0^2\int_0^{x_{i}}\frac{n_{\rm e}(x)^2\langle B_{x}(x)^2\rangle\lambda_{x}}{(1+z(x))^4}\mathrm{d}x\approx\chi_{0}+\displaystyle\sum_{j}l_{ij}\chi_{j}.
\label{diffenv}
\end{equation}
The intrinsic contribution of the emitting source $\chi_0$ is assumed
to be on average the same for all sources.  Since the intrinsic size
of the emitting radio source is unknown, $\chi_0$ is defined according
to Eq.\,(\ref{chidef}) but with the length $l_{i0}$ factored in. Here,
the auto-correlation function of the magnetic field has been
approximated with a top-hat function of width $\lambda_{x}$. We assume
that, within a correlation length $\lambda_x$, the redshift can be
approximated to be constant.  Moreover, we note that $\chi_{j}$ has
been assumed to be independent of the redshift, meaning that the
contribution to the Faraday rotation from a given environment does not
change with redshift. The redshift dependence is totally absorbed by
the term $l_{ij}$.  Residual small-scale, uncorrelated Galactic
contributions could be present in the estimate of the extragalactic
Faraday rotation $\phi_{{\rm e}, i}$. To properly take them into
account, the inclusion of a latitude dependent contribution in
Eq.\,(\ref{diffenv}) could be necessary.

To constrain the contribution from the different environments
$\chi_{j}$, an estimate of the extragalactic Faraday rotation
$\phi_{{\rm e}, i}$ for a collection of lines of sight, as well as a
description of the cosmic web large-scale structure, summarized by
$l_{ij}$, is required. With a Bayesian approach, Jasche et al.\ (2010)
reconstructed the cosmic density field on the basis of optical data
from the SDSS \emph{Data Release 7}.  Following the scheme proposed by
Hahn et al.\ (2007), they give a three-dimensional classification of
the density field posterior in terms of voids, sheets, filaments, and
clusters of galaxies.  For a source with known redshift, this
reconstruction allows one to estimate the path covered by the radio
signal through each environment, i.e., the elements $l_{ij}$ in
Eq.\,(\ref{diffenv}).  Radio sources can sit in different structures,
e.g. galaxy clusters or filaments. This effect can be taken into
account statistically since it is possible to locate the radio source
in the reconstruction of the large scale structure via redshift
identification.  The fractions of the line of sight that the different
radio sources spend in the different environments varies (this is also
the case for the same radio source when different reconstruction of
the large scale structure are considered). The combination of this
information with the Faraday depth estimation for a collection of
sources enables us therefore to statistically discriminate between the
average contributions from the source itself, from galaxy clusters,
filaments, voids, and sheets.

\section{Bayesian approach}
To infer information about the contribution from the different extragalactic
environments, $\chi$, to the extragalactic Faraday rotation on the basis of
data $d$ \footnote{Here, $d$ and $\chi$ are respectively vectors with
  elements $d_{i}$ and $\chi_{j}$ with $i=1,...,N_{\rm los}$ and $j=0,...,
  N_{\rm env}$, where $N_{\rm los}$ and $N_{\rm env}$ are the total number of
  lines of sight and the total number of environments in between the source
  and the observer.}  obtained from a RM grid, we propose a
Bayesian approach. Bayes theorem,
\begin{equation}
P(\chi|d)=\frac{P(d|\chi)P(\chi)}{P(d)},
\label{ourposterior}
\end{equation}
expresses our \emph{a posteriori} knowledge $P(\chi|d)$ about the
signal $\chi$ after the measurement process took place.  The initial
knowledge on the signal before the data have been acquired is
described by the prior $P(\chi)$ and is modified by the data $d$
through the likelihood $P(d|\chi)$. The evidence $P(d)$ is a
normalization factor, obtained by marginalizing the joint probability
$P(d,\chi)=P(d|\chi)P(\chi)$ over all the possible configurations of
the signal $\chi$.

In order to get a data-driven solution and to keep our assumptions as general
as possible, we do not put constraints on the magnetic field strength and
assume all values of magnetic field magnitudes to have equal probability.  Such
\emph{a priori} knowledge can be well described by a flat prior on a
logarithmic scale. Mathematically, it is expressed as an Inverse-Gamma
distribution
\begin{equation}
P(\chi)\propto\displaystyle\prod_{j}\,\chi_{j}^{-\alpha_{j}}e^{-\frac{q_{j}}{\chi_{j}}}
\label{prior}
\end{equation}
where $\alpha_{j}=1$ and $q_{j}=0$ (Jeffreys' Prior). 

The Faraday depths $d_{i}$ coming from radio observations contain the
Galactic and extragalactic contribution, $\phi_{{\rm g},i}$ and
$\phi_{{\rm e},i}$, and the noise $n_{i}$ of the measurement process
\begin{equation}
d_{i}=\phi_{\rm g, i }+\phi_{\rm e, i}+ n_{i}.
\end{equation}
Oppermann et al.\ (2012, 2014) developed an algorithm to reconstruct
the Galactic Faraday rotation foreground and estimate the
extragalactic contribution, given a catalog of Faraday RMs for
different lines of sight.  Their results are available as a collection
of $N$ random samples extracted from the posterior distribution
\begin{equation}
P(\phi_{\rm e}|d, {\rm Opp})=\frac{P(d|\phi_{\rm e})P(\phi_{\rm e}|{\rm Opp})}{P(d|{\rm Opp})},
\label{nielspost}
\end{equation}
which intrinsically depends on the assumption enclosed in their prior,
as indicated by ``Opp''.  In order to disentangle the different
contributions that make up the overall observed extragalactic Faraday
depth, this prior must be replaced with a new one.

If in Eq.\,(\ref{ourposterior}), we expand our likelihood
\begin{equation}
P(d|\chi)=\int \mathcal{D}\phi_{\rm e}P(d|\phi_{\rm e})P(\phi_{\rm e}|\chi)
\label{explike}
\end{equation}
and insert the likelihood $P(d|\phi_{\rm e})$ from Oppermann et al.\ (2014), see Eq.\,(\ref{nielspost}), we obtain
\begin{equation}
P(\chi|d)=\int \mathcal{D}\phi_{\rm e}\frac{P(\phi_{\rm e}|d,
  {\rm Opp})P(d|{\rm Opp})}{P(\phi_{\rm  e}|{\rm Opp})}\frac{P(\phi_{\rm e}|\chi)P(\chi)}{P(d)},
\end{equation}
where $\int \mathcal{D}\phi_{\rm e}$ denotes the phase space integral
over all the possible field configurations of $\phi_{\rm e}$. As the
evidence is simply a normalization factor, this posterior no longer
depends on the assumptions of Oppermann et al.\ (2014), which cancel
out.

Under the assumption that the lines of sight are sufficiently
separated, that is, the cross-correlation function of their magnetic
field is zero, for each line of sight we can consider independent
Gaussian distributions for the extragalactic Faraday depth given the
cosmic web's large-scale structure
\begin{equation}
P(\phi_{\rm e}|\chi)= \int \mathcal{D}l P(l|{\rm Jasche}) \displaystyle\prod_{i}\mathcal{G}(\phi_{\rm e,i}, \displaystyle\sum_{\rm  j}l_{ij}\chi_{j}),
\end{equation}
where $P(l|{\rm Jasche})$ is the posterior of the cosmic web structure from
Jasche et al.\ (2010), available in the form of $N^{\prime}$ samples.  The
notation $\mathcal{G}(x,X)$ indicates a one-dimensional Gaussian distribution
for a variable $x$ with zero mean and variance $X$.  Moreover, since the prior
$P(\phi_{\rm e}|{\rm Opp})$ from Oppermann et al.\ (2014) is
\begin{equation}
P(\phi_{\rm e}|{\rm Opp})=\displaystyle\prod_{i}\mathcal{G}(\phi_{{\rm e},i}, \sigma_{\rm e}^2)
\label{nielsprior}
\end{equation}
with a standard deviation $\sigma_{\rm e}\sim$6.4\,rad/m$^2$, it follows that
\begin{equation}
P(\chi, \phi_{\rm e}, l|d)\approx\frac{\displaystyle\prod_{i}\mathcal{G}(\phi_{{\rm e},i}, \displaystyle\sum_{j}l_{ij}\chi_{j})}{\displaystyle\prod_{i}\mathcal{G}(\phi_{{\rm e},i}, \sigma_{\rm e}^2)}\displaystyle\prod_{j}\chi_{j}^{-1}P(\phi_{\rm  e}|d, {\rm Opp})P(l|{\rm Jasche}).
\end{equation}
This probability density function can be marginalized over $l$ and
$\phi_{\rm e}$ by making use of the sample-representation available
for the posterior of the large scale structure $P(l | {\rm Jasche})$
and of the extragalactic Faraday rotation $P(\phi_{\rm e}|d, {\rm
  Opp})$.  To evaluate the posterior distribution for $\chi$, we plan
to use a Monte Carlo Markov Chain approach.

In this approximation the lines of sight are considered
independent. This means that, if the assumption does not hold, a
correlated component of the magnetic field could be missed.  A proper
modeling that takes into account possible cross-correlations between
lines of sight is left for future works.

\section{Current developments and applications with SKA}
An application to archival \emph{Very Large Array} (VLA) data is in
progress. By reanalyzing 1.4\,GHz data from the \emph{NRAO VLA Sky
  Survey} (NVSS), Taylor et al.\ (2009) derived a catalog of RM values
for 37543 linearly polarized radio sources, north of declination
$-40^{\circ}$ - which corresponds to a number density of approximately
one source per square degree.  Oppermann et al.\ (2014) evaluated the
posterior distribution of the extragalactic Faraday rotation for all
the lines of sight in this catalog.

By using optical data from on-line databases (NED and SIMBAD) as well
as from surveys (SDSS, 6dFGS, 2dFGRS, 2QZ, 6QZ), Hammond et
al.\ (2012) published a catalog of spectroscopic redshifts for 4003
linearly polarized sources from the NVSS. The sources in this sample
have redshifts in the range $0.0007<z<5.3$.  Only about 30 sources
have $z<0.02$, and the mean redshift of the sample is 0.89.  The mean
angular distance of the sources being equal to $\sim$1 degree, this
implies a mean distance among sources larger than 1\,Mpc for 99\% of
the sample, and on average larger than $\approx$30\,Mpc. If
$\lambda_x>>$30\,Mpc, the assumption of no cross-correlation is not
valid anymore. As noted in the previous section, this would translate
in an underestimate of the magnetic field for structures with such a
length scale.  By combining the reconstruction of the large-scale
structure from Jasche et al.\ (2010) with this catalog of redshifts,
it is possible to evaluate the path covered by the radio signal
through each environment for each line of sight.

The catalog of Faraday rotation values in Taylor et al.\ (2009) was
derived through a linear fit of the observed polarization angle versus
$\lambda^2$ at two closely separated radio frequencies (1365 and
1435\,MHz). When few data points are used, such $\lambda^2$-fits can
suffer from an $n\pi$-ambiguity. Moreover, this approach is suitable
only if no Faraday rotating plasma is mixed with the radio-emitting
plasma. A better analysis of polarization properties of radio sources
can be done with the RM synthesis (Brentjens \& de Bruyn 2005) and the
Faraday synthesis (Bell \& En{\ss}lin 2012) techniques if more
frequencies are available. A continuous and broad frequency coverage
is desirable, both to reduce the risk of $n\pi$-ambiguities and to
reach a sufficiently high resolution in Faraday depth to distinguish
nearby Faraday components (e.g., Farnsworth et al.\ 2011, Kumazaki et
al.\ 2014).

In this respect, the \emph{Square Kilometer Array} as well as its
precursors will offer a great improvement.  Broadband
spectro-polarimetric surveys of the sky in combination with deep
observations of specific targets will provide densely spaced RM grids.
A RM catalog one hundred times denser than those currently existing
(Taylor et al.\ 2009) will be obtained with POSSUM, the polarization
survey planned with the SKA precursor ASKAP. POSSUM will observe the
sky between 1130 and 1430\,MHz with a sensitivity in U and Q of
$\sim10$\,$\mu$Jy/beam and a resolution of 10$^{\prime\prime}$, giving
an expected density of 70 polarized sources per square degree (Hales
et al.\ 2014a).  The uncertainty on the RM is lower than $\approx
6$\,rad/m$^2$ for a polarized signal with a $S/N>5$ (Stepanov et
al. 2008).  The ultra-deep polarization survey CHILES Con Pol with the
Jansky Very Large Array (Hales et al.\ 2014b) will reach a sensitivity
of 400\,nJy with sky area of a fraction of a square degree, providing
a first look at the magnetic field science that can be performed with
the SKA.  The sky survey in polarization SKA1-SUR planned for SKA1
will go much further. Designed to cover the frequency range
650-1670\,MHz, it will produce Stokes U and Q images over 30000 square
degrees with a spatial resolution of 2$^{\prime\prime}$. 
Within two years, a
sensitivity in polarization of $\sim1\mu$Jy/beam will be reached, with
an uncertainty in RM lower than $\approx 1$\,rad/m$^2$ ($S/N>5$).  The
high RM accuracy will make our approach viable for detecting magnetic
fields with strengths $\sim$\,nG already during the early science
phase of SKA1, when sensitivity will be at 50\% of its full level. For
weaker magnetic fields the full deployment of all the SKA1 baseline will
be necessary.  The resulting sky grid of Faraday rotation values will
be 300-1000 times denser than the largest catalog available at the
moment. The better resolution of 2$^{\prime\prime}$, compared to the
45$^{\prime\prime}$ of Taylor et al.\ (2009), will make it possible to
identify optical counterparts uniquely and hence the source redshift.
For fields of particular interest, e.g., regions where galaxy
filaments have been observed, deeper observations could be necessary
to assess the presence of magnetic fields and their properties.  In
this case, the observations from the survey can be complemented by
targeted SKA1-MID pointings over a continuous frequency band from 350
to 3050\,MHz.  This would imply a polarization sensitivity
$\lesssim1\,\mu$Jy/beam after one hour of integration time and an
uncertainty in RM lower than $\approx 0.2$\,rad/m$^2$ ($S/N>5$).
These targeted observations will be crucial, both to increase the
resolution in Faraday depth and to detect a larger number of sources,
allowing to distinguish nearby components in Faraday spectra and
improve the statistics.

The even higher spatial and frequency resolution and the deeper
sensitivity of SKA2 will make it possible to obtain RMs for 
an unprecedented
number of background and intervening sources. It is
therefore essential to develop and apply 
statistical approaches with the aim of interpreting the data correctly.

\section{Conclusions}
The origin and evolution of cosmological magnetic fields are still
open questions. Magnetic fields are currently being studied in galaxy
clusters but a firm detection is lacking in larger scale structures,
such as galaxy filaments.  The structure and the strength of magnetic
fields in galaxy clusters are both deeply affected by the cluster
formation processes (e.g., Dolag et al.\ 2005, Donnert et al.\ 2009,
Xu et al.\ 2010), almost independently of the mechanism of generation
of the seed magnetic field. This makes it difficult to discriminate
between the different formation scenarios from galaxy cluster
observations alone.  The SKA's large bandwidth, high sensitivity and
frequency resolution will permit a more reliable recovery of the
Faraday depth for a large number of sources, and its high spatial
resolution will allow a proper identification of the optical
counterparts necessary for redshift assignment. Modern techniques
based on the largest photometric all-sky surveys (2MASS, WISE and
SuperCOSMOS) already provide catalogs of redshifts for millions of
sources up to redshifts of 0.2 to 0.3 (e.g. Bilicki et al. 2014), and
the next generation of redshift surveys will push those limits. We
will investigate to what depths and accuracy redshifts are needed in
order to apply the proposed method. The ultimate goal is to search for
magnetic fields in large-scale structures like filaments, where
magnetic field amplification is not yet saturated and therefore the
field strength still depends on the seed field strength.  This
information will help to place firm constraints on the origin and
evolution of cosmological magnetic fields.

\acknowledgments We thank the anonymous referee for useful comments
and suggestions. The implementation of the code makes use of the NIFTY
package by Selig et al.\ (2013) and of the cosmology calculator by Ned
Wright (www.astro.ucla.edu/$\sim$wright).  This research was supported
by the DFG Forschengruppe 1254 ``Magnetisation of Interstellar and
Intergalactic Media: The Prospects of Low-Frequency Radio
Observations''.

\end{document}